# CMEs during the Two Activity Peaks in Cycle 24 and their Space Weather Consequences

N. Gopalswamy[1], P. Mäkelä[1,2], S. Akiyama[1,2], S. Yashiro[1,2], and N. Thakur[1,2]

[1]NASA Goddard Space Flight Center, Greenbelt, Maryland, USA

[2]The Catholic University of America, Washington DC, USA

E-mail: nat.gopalswamy@nasa.gov

**Abstract**

We report on a comparison between space weather events that occurred around the two peaks in the sunspot number (SSN) during solar cycle 24. The two SSN peaks occurred in the years 2012 and 2014. Even though SSN was larger during the second peak, we find that there were more space weather events during the first peak. The space weather events we considered are large solar energetic particle (SEP) events and major geomagnetic storms associated with coronal mass ejections (CMEs). We also considered interplanetary type II radio bursts, which are indicative of energetic CMEs driving shocks. When we compared the CME properties between the two SSN peaks, we find that more energetic CMEs occurred during the 2012 peak. In particular, we find that CMEs accompanying IP type II bursts had an average speed of 1543 km/s during the 2012 peak compared to 1201 km/s during the 2014 peak. This result is consistent with the reduction in the average speed of the general population of CMEs during the second peak. All SEP events were associated with the interplanetary type II bursts, which are better than halo CMEs as indicators of space weather. The comparison between the two peaks also revealed that the discordant behavior between the CME rate and SSN was more pronounced during the second peak. None of the 14 disk-center halo CMEs was associated with a major storm in 2014. The lone major storm in 2014 was due to the intensification of the (southward) magnetic field in the associated magnetic cloud by a shock that caught up and propagated into the magnetic cloud.

**Key words:** coronal mass ejections, Sunspot number, solar activity, space weather, solar energetic particle events, geomagnetic storms

## 1. Introduction

Solar cycle 24 has been extremely weak as measured by the sunspot number (SSN) and is the smallest since the beginning of the Space Age. The weak activity has been thought to be due to the weak polar field strength in cycle 23. Several authors have suggested that the decline in cycle 24 activity might lead to a global minimum (see e.g., Padmanabhan et al., 2015; Zolotova and Ponyavin, 2014). The weak solar activity has been felt throughout the heliosphere, with diminished solar wind speed, density, and magnetic field (McComas et al., 2013; Gopalswamy et al., 2014a,b). On the other hand, the rate of coronal mass ejection (CME) occurrence has not diminished as much, which is not fully understood (Petrie, 2013; Wang and Colaninno, 2013; Gopalswamy et al., 2015a). The space weather in cycle 24 has been extremely mild even with the high rate of occurrence of CMEs. In particular, the numbers of major geomagnetic storms (Dst ≤-100 nT) and high-energy solar energetic particle (SEP) events (>500 MeV) have been very infrequent (Gopalswamy et al., 2014a,b). The cause of the weak geomagnetic storms has been traced to the anomalous expansion of CMEs due to the reduced total pressure in the heliosphere (Gopalswamy et al., 2014a). The reduced magnetic field in the heliosphere has been suggested one of the reasons for the lack of high-energy SEP events because the particle-acceleration efficiency of a CME-driven shock is proportional to the ambient magnetic field strength (see e.g., Kirk, 1994). In addition to the cycle-to-cycle variability, there is additional variability due to the asymmetric activity between the two hemispheres.

It is well known that most solar cycles show a double peak due to the out-of-phase activity in the two hemispheres. The double peak in SSN during cycle 24 is unusual in that the second peak is larger than the first one by ~20%. Such a behavior was observed only a few times since the 1800s (Gopalswamy et al., 2015a). Therefore, it is of interest to study the behavior of CMEs during the second peak in solar activity and compare it with the first in order to understand the space weather events of different intensity during the SSN peaks.

In this paper, we investigate the large SEP events and major geomagnetic storms during the two SSN peaks in cycle 24. We also compare the activity in cycles 23 and 24 to provide context to the SSN variability. Since severe space weather is caused by energetic CMEs, we also compare halo CMEs and fast and wide (FW) CMEs during the two peaks. In particular, we consider halo CMEs originating from close the disk center for large geomagnetic storms and halos originating from the western hemisphere for SEP events. We also consider CMEs associated with IP type II bursts, which are indicative of shock-driving CMEs.

## 2. Observations

In order to compare various signatures of solar activity around the two SSN peaks, we consider the daily rate of CMEs averaged over Carrington rotation periods. We use the CME data available online at the CDAW Data Center (cdaw.gsfc.nasa.gov, Gopalswamy et al., 2009). The CME list has been compiled from the images obtained by the Large Angle and Spectrometric Coronagraph (LASCO, Brueckner et al., 1995) on board the Solar and Heliospheric Observatory (SOHO) mission. Since halo CMEs are one of the indicators of energetic CMEs, we make use of the halo CME catalog available at the CDAW Data Center (http://cdaw.gsfc.nasa.gov/CME_list/halo/halo.html, Gopalswamy et al., 2010). Another source of information on shock-driving CMEs is the list of type II radio bursts observed in the decameter-hectometric (DH) and kilometric (km) wavelengths by the Radio and Plasma Wave Experiment (WAVES, Bougeret et al., 1995) on board the Wind spacecraft. A list of these type II bursts along with the solar sources and the associated CMEs is also available at the CDAW Data Center (http://cdaw.gsfc.nasa.gov/CME_list/radio/waves_type2.html). Coronagraph and WAVES data from the Solar Terrestrial Relations Observatory (STEREO, Howard et al., 2008) mission are also used to cross check source locations of CMEs and the wavelength range of type II bursts. Finally, we use the list of large SEP events (Gopalswamy et al., 2015b) and major geomagnetic storms (Gopalswamy et al., 2015c) available in the literature to compare the space weather events around the two SSN peaks. A list of large SEP is also available at the CDAW Data Center (http://cdaw.gsfc.nasa.gov/CME_list/sepe/). This list includes the associated CMEs and their solar sources. We take the events in the years 2012 and 2014 as representative of the first and second SSN peaks, respectively.

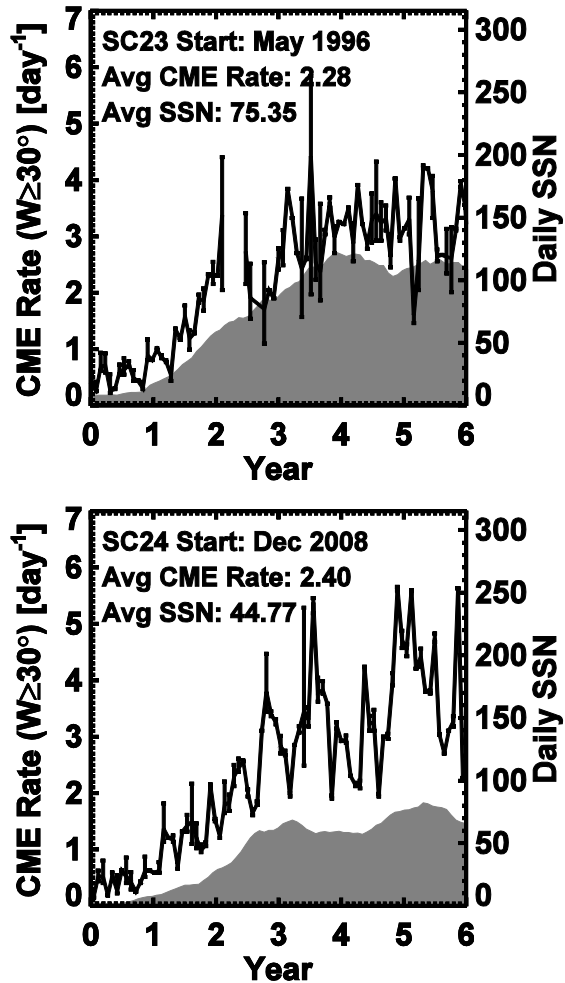

Fig.1. Daily rate of CMEs (averaged over the Carrington rotation period of 27.24 days) compared with the daily international sunspot number (SSN) obtained from Solar Influences Data Center (http://sidc.oma.be/sunspot-data/). CMEs with width ≥30° are included. The averages over the first 73 months of the two cycles are shown in the plot. The error bars are obtained based on LASCO data gaps that are >3 h.

## 3. Sunspot Number and CME Occurrence Rate

Figure 1 shows the daily CME rate averaged over the Carrington rotation period, along with SSN for the first six years of cycles 23 and 24. We see that the interval between the two SSN peaks was larger in cycle 24 (~2 years) than in cycle 23 (~1.5 year). Furthermore, the second SSN peak was more pronounced in cycle 24, which is opposite to that in cycle 23. We have considered only CMEs with width ≥30° to avoid coronal changes and ill-defined CMEs. We see that there was an overall increase in SSN and CME rate toward the maximum phase, but the difference between the two phenomena is conspicuous in cycle 24. The CME rate variability in cycle 24 is generally higher than that in cycle 23. Even though the SSN (averaged over the first 73 months of each cycle) has dropped by ~40% from ~76 to ~45, the CME rate remained about the same. When normalized to SSN, the CME rates become 0.05/SSN in cycle 24 vs. 0.03/SSN in cycle 23, showing that the discordance between SSN and CME rate increased in cycle 24.

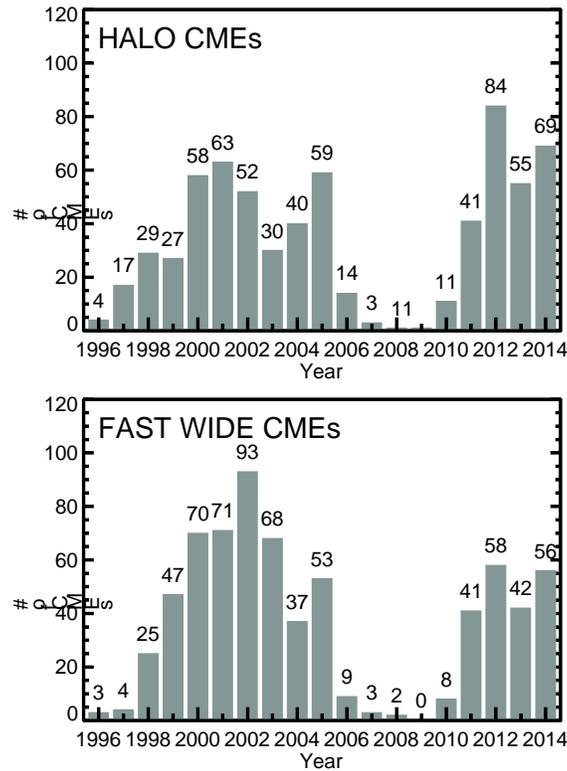

Fig.2. Annual rates of halo CMEs (top) and fast and wide CMEs (bottom) from 1996 to the end of 2014. Cycle 23 started in May 1996. Cycle 24 started in December 2008. Halo CMEs are those that appear to fully encircle the occulting disk of the LASCO/C3 coronagraph. CMEs with speeds ≥900 km/s and width ≥60° are considered to be fast and wide (FW). The number of CMEs in each bin is given on the plots. Note that the double-peak structure is not observed in cycle 23, while it is clear in cycle 24.

Figure 2 shows the annual rates of halo CMEs and FW CMEs. These are special CME populations most relevant for space weather applications. The annual numbers of halo CMEs in the three maximum years in cycle 23 were similar or slightly smaller compared to those in cycle 24. On the other hand, the number of FW CMEs over the same period was larger in cycle 23. When normalized to SSN, the FW CME rates were similar in the two cycles. Halo CMEs and FW CMEs did not show the double peak in their annual numbers in cycle 23. The substantial difference between the two SSN peaks in cycle 24 was also reflected in the annual number of halo CMEs and the number of FW CMEs. Figure 2 shows that the annual number was higher (84) during the first peak than that during the second peak (69). Similarly, the number of FW CMEs was higher during the first peak (58 vs. 52 in the second peak). We expect a similar difference in space weather events between the peaks because both large SEP events and major geomagnetic storms are caused by energetic CMEs.

Table 1. Summary of solar activity in 2012 and 2014

|  | 2012 | 2014 |
|---|---|---|
| Peak SSN | 67 | 90 |
| #Halo CMEs | 84 | 63 |
| #DC Halos | 17 | 14 |
| #Western Halos | 21 | 10 |
| #FW CMEs | 58 | 52 |
| #LSEP Events | 15 | 7 |
| #Major storms | 6 | 1 |
| #DH-km Type II | 19 | 16 |
| DC Halo <V>[a] | 975 km/s | 753 km/s |
| Western Halo <V> | 1088 km/s | 781 km/s |
| DH-km <V> | 1543 km/s | 1201 km/s |

[a]<V> denotes average speed

Table 1 summarizes the activity around the two SSN peaks that are relevant for comparing space weather events. Table 1 lists the peak SSN, the number of halo CMEs, disk-center (DC) halos (those originating within 30° from the disk center), FW CMEs, large SEP events (events with proton intensity ≥10 pfu in the GOES >10 MeV channel; the particle flux unit is defined as 1 pfu = 1 particle cm$^{-2}$ s$^{-1}$ sr$^{-1}$), major geomagnetic storms (Dst ≤-100 nT), and interplanetary (IP) type II bursts in the DH-km range. The DC halo CMEs were used to assess the geoeffectiveness (ability to cause major geomagnetic storms) of the CMEs in 2012 and 2014. Similarly, the western hemisphere halos were used to assess the ability of CMEs in accelerating earth-arriving SEPs. Finally, we considered the DH-km type II bursts, which are indicators of shock-driving CMEs that might also accelerate SEPs. In Table 1, we have given the average sky-plane speeds of the three populations of CMEs during the two SSN peaks. We analyze these numbers in the next section to understand the difference between the two activity peaks.

## 4. Space Weather Events around the Two SSN Peaks

The number of space weather events during the first SSN peak was substantially higher than that during the second peak as can be seen in Table 1. There were 6 major geomagnetic storms in 2012, compared to just one in 2014. There were 15 large SEP (LSEP) events in 2012 compared to just 7 in 2014. While the presence of FW CMEs is a common requirement for both major storms and LSEP events, other requirements are different. For example, storm-causing CMEs need to originate close to the disk center and possess southward magnetic field component either in the CME main body or in the shock sheath ahead of the CME. On the other hand, SEP-associated CMEs need to drive a strong shock irrespective of the internal structure and the CMEs need to originate from the western hemisphere for good magnetic connectivity to Earth (so the particles can be detected near Earth at 1 AU).

### 4.1 CMEs and Geomagnetic Storms

From Table 1 we see that only ~20% of the halos originated within 30° from the disk center during both peaks. Thus the opportunity for CMEs impacting Earth was substantially reduced, but this is true for both peaks. The average speed of DC halo CMEs was ~23% lower during the second SSN peak (753 km/s vs. 975 km/s). Note that this difference was larger than the typical error (~10%) in CME height-time measurements used in obtaining the average speeds. This suggests that the geoeffectiveness of DC CMEs is expected to be smaller in 2014, consistent with the number of storms in Table 1. In order to compare the speeds of the DC halos with those of all storm-causing CMEs, we have shown the speed distributions of such CMEs in cycles 23 and 24 in Figure 3. There were only 11 CMEs that caused major storms during cycle 24 until the end of 2014 (see Gopalswamy et al., 2015b). The average speeds of such CMEs were 838 km/s and 968 km/s for cycle 23 and 24, respectively. From Table 1, we see that the average speed of the DC halos during the first SSN peak was 975 km/s, which is close to the average speed of 968 km/s in Figure 3. On the other hand, the average speed during the second peak (753 km/s) was 22% below the average speed in cycle 24.

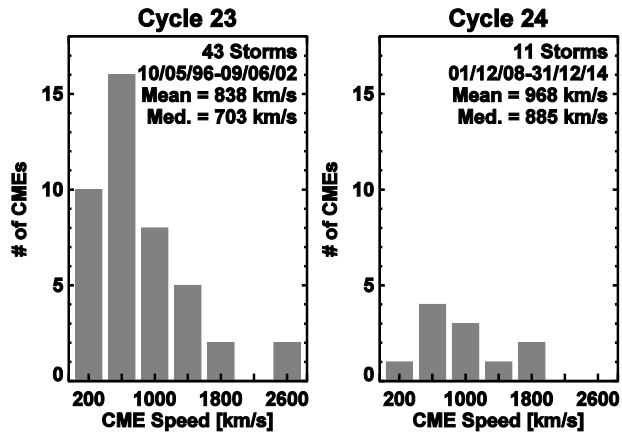

Fig. 3. Distribution of CME speeds associated with major geomagnetic storms of cycle 23 and 24 over the first 73 months in each cycle. The speeds are from the SOHO/LASCO CME catalog (http://cdaw.gsfc.nasa.gov). The average speed in cycle 24 is higher than that in cycle 23, but the number of storms is very small in cycle 24.

None of the 14 DC halos (Table 1) during the second peak in cycle 24 resulted in a major storm. The lone major storm in 2014 was due to the interaction of a CME-shock with a preceding CME (to be discussed below). During the first peak, only two of the 17 DC halos were associated with a major storm (2012 March 07 and July 12 CMEs). One of the remaining 4 major storms was due to a halo CME on 2012 September 28 not too far from the disk center (N06W34). The three remaining storms were due to non-halo CMEs (2012 April 19 at 15:12 UT from S30E71, 2012 October 5 at 02:48 UT from S23W31 and 2012 November 09 at 15:12 UT from S18E16), so they are not included in Table 1. The three CMEs were partial halos as the widths were in the range 142º to 284º. The major storm on April 24 has been tentatively associated with the 2012 April 19 CME. The CME was associated with a filament eruption from S30E71. This CMEs erupted far from the central meridian, although the CME expanded westward during propagation. There were also a couple of smaller and slower CMEs from the western hemisphere on the same day. Therefore, there is some uncertainty in the source identification of this storm. However, solar wind observations clearly indicate an IP CME (ICME) arriving at Sun-Earth L1 on April 23 at 16:35 UT preceded by a shock at 02:30 UT on the same day (see Figure 4). A number of solar wind parameters (http://omniweb.gsfc.nasa.gov/) are plotted in Figure 4: the magnitude of the IP magnetic field (IMF) strength ($B_t$), the east-west ($B_y$) and north-south ($B_z$) components of the IMF, solar wind speed (V), density (N), temperature (T), plasma beta and the Dst index. The Dst index reached its minimum value (-108 nT) at 5:00 UT on April 24. There was a stream interaction region and a high speed solar wind stream immediately following the ICME. These might help in narrowing down the responsible CME because the coronal hole needs to be located to the east of the CME source. The magnitude of the southward component of the CME magnetic field in the IP medium was ~15 nT, so the storms was not very intense. The field magnitude was relatively low during most of the storms in cycle 24 (Gopalswamy et al., 2015c), attributed to the increased CME expansion in cycle 24 (Gopalswamy et al., 2014a).

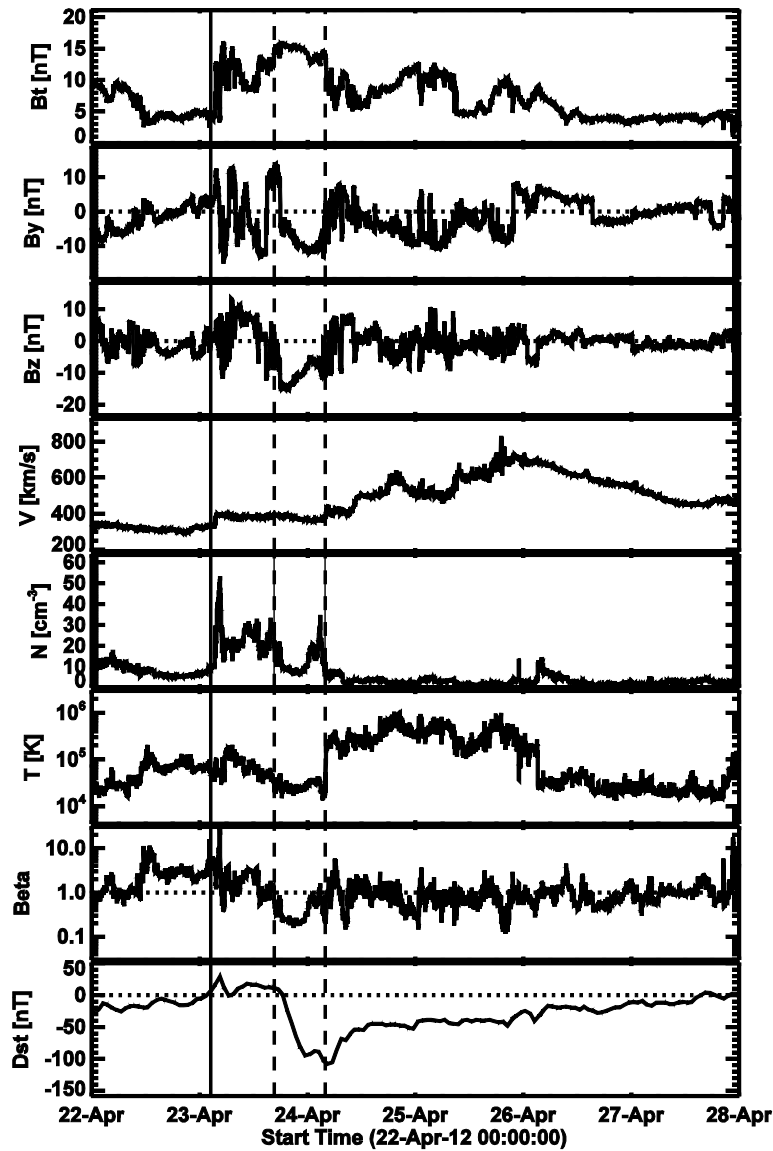

Fig. 4. Time profiles of plasma parameters, magnetic field components (OMNI 1-minute data) and the Kyoto Dst index (1-hour time resolution) during 22-28 April 2012. From top to bottom: the total magnetic field strength ($B_t$), Y- and Z- components of the magnetic field ($B_y$, $B_z$), solar wind flow speed (V), proton density (N), proton temperature (T), plasma beta, and Dst index. The low plasma beta values bounded by the vertical dashed lines indicate the approximate boundaries of the ICME (23/16:35 UT – 24/03:55 UT). The sharp increase in N and T near the rear ICME boundary indicates the stream interaction region.

We now discuss the lone major storm during the second SSN peak in cycle 24 that was due to the passage of a shock into a preceding CME. Figure 5 plots $B_t$, $B_y$, $B_z$, V, N, T, plasma beta and the Dst index. The Dst can be seen reaching a minimum value of -60 nT and then recovering to -50 nT before dipping again and reaching the level of a major storm (-112 nT) at 9:00 UT on 2014 February 19. This is a double-dip storm, where the first dip is not due to the shock sheath as in the classical case (Kamide et al., 1998; Gopalswamy, 2009), but both dips are due to the same $B_z$ structure without a northward turning. The dividing line between the two dips in this event is the IP shock that passes through a preceding magnetic cloud (MC). Thus, the sheath of the shock was the rear part of the preceding, fully south-pointing magnetic cloud. The preceding

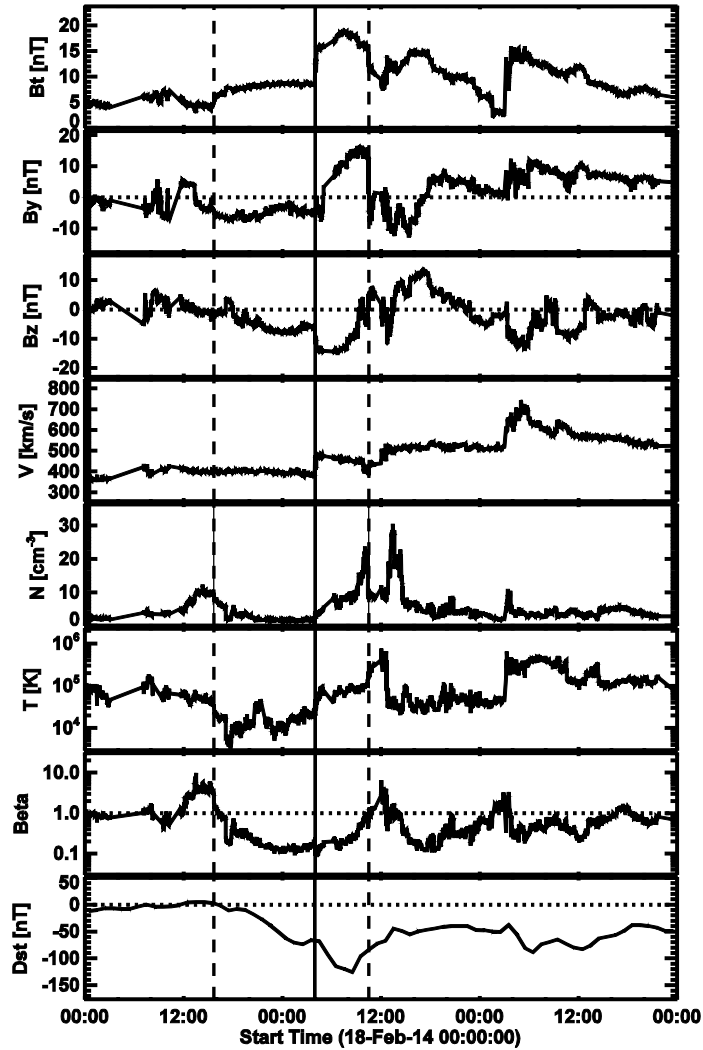

Fig. 5. Time profiles of IMF components (OMNI 1-minute data), plasma parameters, and the Kyoto Dst index (1-hour time resolution) during 18-20 February 2014. From top to bottom: the total magnetic field strength ($B_t$), Y- and Z- components of the magnetic field ($B_y$, $B_z$), solar wind flow speed (V), proton density (N), proton temperature (T), plasma beta, and Dst index. The low plasma beta values bounded by the vertical dashed lines indicate the approximate boundaries of the magnetic cloud (MC, 18/15:40 UT – 19/10:30 UT). The southward $B_z$ of the MC was enhanced by the shock (indicated by the vertical solid line; ~19/03:57 UT) driven by the following halo CME first observed by SOHO/LASCO at 10:00 UT, 16 February 2014.

MC had a minimum $B_z$ of -8 nT. When the shock entered the cloud, the $B_z$ decreased to -14 nT. We can estimate the minimum Dst from the empirical formula, Dst = 0.01 $VB_z$ – 32 nT (Gopalswamy, 2010) using the sheath speed at the time of the shock-cloud interaction (~500 km/s), and $B_z$ = -14 nT, to be -102 nT, similar to the observed Dst value (-112 nT). If there was no cloud-shock interaction, the minimum Dst would not have attained -112 nT. If we use $B_z$ = - 8 nT and V = 400 km/s, the Dst would have attained a minimum value of -64 nT (moderate storm level) according to the above formula. This value is once again close to the observed value corresponding to the first dip (-60 nT). We conclude that the only major storm in 2014 was due to the interaction of a shock with a preceding CME; otherwise it would have been only a moderate storm. The CME driving the shock was not geoeffective by itself because its axis was north pointing. Shocks propagating through preceding CMEs have been studied only occasionally (Collier et al., 2007; Echer et al., 2010; Lugaz et al., 2015). Collier et al. (2007) showed that about 10% of magnetic clouds may have shocks propagating through them. Lugaz et al. (2015) also found a similar percentage among events in solar cycle 23. Several events in the list published by Lugaz et al. (2015) show a Dst change similar to what was observed during the 2014 February 19 storm.

## 4.2 CMEs and SEP Events

Table 1 shows that the number of halo CMEs from the western hemisphere in 2012 was two times larger than that in 2014, similar to the number of LSEP events. Furthermore, the average speed of the western hemisphere halos in 2012 (1088 km/s) was higher than that in 2014 (781 km/s) by ~40%. This is also consistent with the different speeds of the DC halos during the two SSN peaks. Among the 15 LSEP events in 2012, only 8 had overlap with the 21 western halos. The remaining 7 were associated with eastern halos or behind the west limb events. The halos that were not associated with LSEP events were generally very slow (317 to 917 km/s). Some of the halos with intermediate speeds were associated with minor SEP events (proton intensity <10 pfu) or they occurred during elevated background SEP intensity, so it is hard to decide the SEP association. The situation was similar in 2014 with only 3 of the 7 LSEP events overlapping with the 10 western halos. The halos without SEP association were generally slow (the average speed was 577 km/s in 2012 and 476 km/s in 2014). The average speed of the SEP-associated halos during the first peak was 1680 km/s, while it was 1327 km/s in 2014. Clearly, the CME speed is an important factor that decides whether CMEs are associated with an SEP event or not.

## 4.3 CMEs and Interplanetary Type II Bursts

Type II radio bursts extending over a wide wavelength range are indicative of strong shocks in the IP medium (Gopalswamy et al., 2005). Of particular interest are the bursts that have emission components in the DH and km wavelengths. DH-km type II bursts are highly associated with SEP events because the same shock accelerates electrons to produce type II bursts and ions observed as SEP events. The type II association with SEP events at Earth depends on the source location of type II bursts. It has been shown that DH type II bursts originating from the western hemisphere of the Sun have high degree of SEP association (Gopalswamy et al., 2008). Tables 2 and 3 list the DH-km type II bursts observed by the Wind/WAVES experiment in 2012 and 2014, respectively. The starting frequency of the type II bursts (fs in MHz) is limited by the upper frequency cutoff of the Wind/WAVES (14 MHz) or

STEREO/WAVES (16 MHz) instruments. The bursts may or may not have a metric type II component, which we ignore for the present study. We see that the number of DH-km type II bursts was similar in 2012 (19) and 2014 (16). Most of the CMEs associated with type II bursts were halos in 2012 (16 out of 19 or 84%) and 2014 (15 out of 16 or 94%). The sky-plane speeds of the CMEs were generally high, the lowest speed being 947 km/s in 2012 and 528 km/s in 2014. The average CME speed was clearly higher in 2012 (1543 km/s) than in 2014 (1201 km/s).

Table 2. List of DH-km type II bursts in 2012, the associated CMEs and SEP events

| Type II Date | Type II UT | fs MHz | CME Date | CME UT | CME Width[a] | CME Speed[b] | CME Source[c] | LSEP?[d] |
|---|---|---|---|---|---|---|---|---|
| 2012/01/19 | 15:00 | 16 | 01/19 | 14:36 | H | 1120 | N32E22 | Y |
| 2012/01/23 | 04:00 | 16 | 01/23 | 04:00 | H | 2175 | N28E21 | Y |
| 2012/01/27 | 18:30 | 16 | 01/27 | 18:27 | H | 2508 | N27W71 | Y |
| 2012/03/05 | 04:00 | 16 | 03/05 | 04:00 | H | 1531 | N17E52 | HiB |
| 2012/03/07 | 01:00 | 16 | 03/07 | 00:24 | H | 2684 | N17E27 | Y |
| 2012/03/10 | 17:55 | 14 | 03/10 | 18:00 | H | 1296 | N17W24 | HiB |
| 2012/03/13 | 17:35 | 16 | 03/13 | 17:36 | H | 1884 | N17W66 | Y |
| 2012/03/18 | 00:20 | 16 | 03/18 | 00:24 | H | 1210 | N18W116 | N |
| 2012/03/24 | 00:40 | 16 | 03/24 | 00:24 | H | 1152 | N18E164 | N |
| 2012/05/17 | 01:40 | 16 | 05/17 | 01:48 | H | 1582 | N11W76 | Y |
| 2012/07/05 | 22:40 | 3 | 07/05 | 22:00 | 94 | 980 | S12W46 | N |
| 2012/07/06 | 23:10 | 16 | 07/06 | 23:24 | H | 1828 | S13W59 | Y |
| 2012/07/08 | 16:35 | 16 | 07/08 | 16:54 | 212 | 1495 | S17W74 | Y |
| 2012/07/12 | 16:45 | 14 | 07/12 | 16:48 | H | 885 | S15W01 | Y |
| 2012/07/17 | 14:40 | 12 | 07/17 | 13:48 | 255 | 958 | S28W65 | Y |
| 2012/07/19 | 05:30 | 5 | 07/19 | 05:24 | H | 1631 | S13W88 | Y |
| 2012/07/23 | 02:30 | 16 | 07/23 | 02:36 | H | 2003 | S17W132 | Y |
| 2012/08/31 | 20:00 | 16 | 08/31 | 20:00 | H | 1442 | S25E59 | Y |
| 2012/09/27 | 23:55 | 16 | 09/28 | 00:12 | H | 947 | N06W34 | Y |

[a]Width in degrees (H – halo CME); [b]Speed in km/s; [c]Source location in heliographic coordinates; [d]Indication of an LSEP event (m – minor event, Y – yes, N – no, HiB – high background SEP intensity)

What is remarkable is that all but one of the 15 LSEP events in 2012 were associated with DH-km type II bursts listed in Table 2. The frequency extent of the type II burst for the 2012 May 27 LSEP event was not clear, so we did not include the burst in Table 2. The non-SEP DH-km bursts belonged to one of the following three groups: (1) eastern events, which were not well connected to an Earth observer, (2) there was a high SEP background from earlier events, and (3) CME speeds were low. In 2014, the result is similar: 5 of the 6 LSEP events were associated with DH-km bursts listed in Table 3. The 6th SEP event that occurred on 2014 November 1 was associated with a filament eruption event (Gopalswamy et al., 2015b). This event is not in Table 3 because there was a data gap in Wind/WAVES and STEREO/WAVES observations. There was one DH-km type II in 2014 with a high SEP background (HiB). There were 6 minor SEP events (>10 MeV proton intensity <10 pfu). The large number of minor SEP events is consistent with the lower-speed CMEs during the second peak because of the well-known correlation between CME speed and SEP intensity (e.g., Kahler, 2001). All the non-SEP type II bursts had eastern sources (mostly behind the east limb) so the lack of SEP association may be due to poor connectivity. The average speed of CMEs associated with the DH-km type II burst is a clear distinguishing characteristic between the two SSN peaks in cycle 24.

Table 3. List of DH-km type II bursts in 2014, the associated CMEs and SEP events

| Type II Date | Type II UT | fs MHz | CME Date | CME UT | CME Width[a] | CME Speed[b] | CME Source[c] | LSEP?[d] |
|---|---|---|---|---|---|---|---|---|
| 2014/01/04 | 19:03 | 6.5 | 01/04 | 21:22 | H | 977 | S11E34 | M |
| 2014/01/06 | 07:57 | 14 | 01/06 | 08:00 | H | 1402 | S15W112 | Y |
| 2014/01/07 | 18:33 | 14 | 01/07 | 18:24 | H | 1830 | S15W11 | Y |
| 2014/02/18 | 02:15 | 2.1 | 02/18 | 01:36 | H | 779 | S24W34 | M |
| 2014/02/25 | 00:56 | 14 | 02/25 | 01:25 | H | 2147 | S12E82 | Y |
| 2014/03/25 | 07:52 | 1.7 | 03/25 | 05:36 | 261 | 651 | S14W27 | M |
| 2014/03/29 | 18:00 | 14 | 03/29 | 18:12 | H | 528 | N11W32 | M |
| 2014/04/02 | 13:42 | 14 | 04/02 | 13:36 | H | 1471 | N11E53 | N |
| 2014/04/18 | 13:06 | 14 | 04/18 | 13:25 | H | 1203 | S20W34 | Y |
| 2014/05/07 | 16:24 | 6.3 | 05/07 | 16:24 | H | 923 | S11W100 | M |
| 2014/05/09 | 02:40 | 12 | 05/09 | 02:48 | H | 1099 | S11W122 | M |
| 2014/08/28 | 18:42 | 2.2 | 08/28 | 17:24 | H | 766 | S19E162 | N |
| 2014/09/01 | 11:38 | 6.6 | 09/01 | 11:12 | H | 1901 | N14E127 | N |
| 2014/09/09 | 00:05 | 11 | 09/09 | 00:06 | H | 920 | N12E29 | HiB |
| 2014/09/10 | 17:45 | 14 | 09/10 | 18:00 | H | 1267 | N14E02 | Y |
| 2014/09/24 | 21:02 | 14 | 09/24 | 21:30 | H | 1350 | N13E179 | N |

[a]Width in degrees (H – halo CME); [b]Speed in km/s; [c]Source location in heliographic coordinates; [d]Indication of an LSEP event (m – minor event, Y – yes, N – no, HiB – high background SEP intensity)

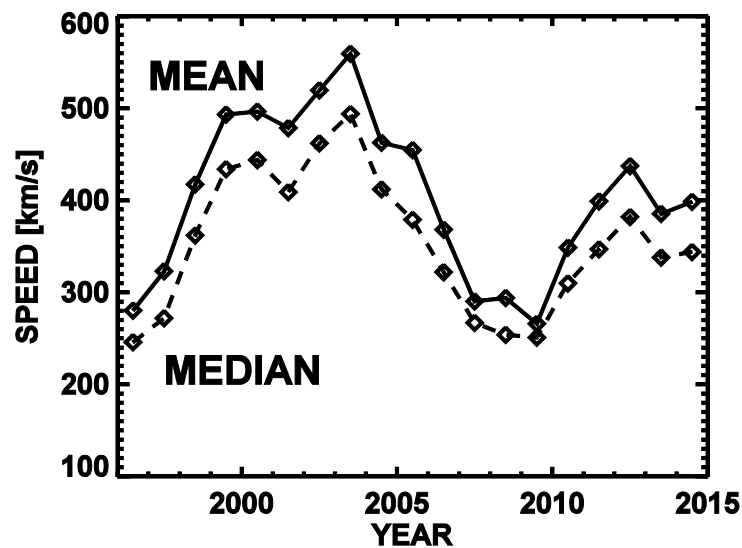

Fig. 6. The mean (upper curve) and median (lower curve) speeds of all CMEs averaged over 12 month periods. CMEs with quality index ≥1 were used to avoid streamer changes and ill-defined CMEs. The speed plots also show the double peak during the maximum phases of cycle 2 and 24. The first peak is dominant in cycle 24 but the second peak is dominant in cycle 23.

The CME speed difference between the two SSN peaks was also observed in the annual averages of CME speeds in general. Figure 6 shows the mean and median speeds of the general population of CMEs from the beginning of SOHO observations in 1996 to the end of 2014, covering cycles 23 and 24. These CMEs have a quality index ≥1 (details can be found in http://cdaw.gsfc.nasa.gov/CME_list/catalog_description.htm). CME-like features with quality index <1 are generally ill-defined, so they are excluded. We see the double peak in the CME speed in both cycles, but the second peak is dominant in cycle 23, while the first peak is dominant in cycle 24. By contrast, the behavior of SSN is opposite: the first peak is dominant in cycle 23 and the second peak in cycle 24 (see Figure 1). While the behavior of SSN is related to the solar dynamo, the behavior of the CME rate and speed depend on how energy is stored and released in solar magnetic regions. Further investigation is needed to understand the discordant behavior of SSN and CME properties.

## 5. Discussion and Summary

The primary result of this investigation is that SSN and CME rates behaved differently around the two peaks of solar activity during solar cycle 24. The second SSN peak (2014) was more pronounced than the first one (2012). However, the number and average speed of halo CMEs were higher during the first peak. Accordingly, the number of space weather events (major geomagnetic storms and LSEP events) is significantly higher during the first peak. The different behavior of SSN and CME rate was also noted in cycle 23 (Gopalswamy, 2004): the SSN was dominant during the first peak, while the CME rate was dominant during the second peak. We used DC halos to assess the geoeffectiveness of the CMEs and western halos for the ability of CMEs to produce Earth-arriving SEPs. It turned out that these halos were not good indicators of space weather events, although they indicated the weakness of CME activity during the second SSN peak. However, the IP type II bursts turned out to be excellent indicators of large SEP events. Almost all SEP events were identified with type II bursts during both the SSN peaks. All SEP events were associated with type II bursts originating from the western hemisphere. The average speeds of CMEs associated with DH-km type II bursts were consistently high but differed substantially between the two SSN peaks (1543 km/s in 2012 vs. 1201 km/s in 2014). Thus the lack of high-energy CMEs seems to be the primary reason for the mild space weather during the second SSN peak.

**Acknowledgments**  This work benefited by NASA's open data policy. We thank the SOHO, STEREO, and Wind teams for making their data available on line.  This work was presented at the UN/Japan Workshop on ISWI supported in part by JSPS Core-to-Core Program (B. Asia-Africa Science Platforms), Formation of Preliminary Center for Capacity Building for Space Weather Research and International Exchange Program of National Institute of Information and Communication Technology (NICT). Work supported by NASA's LWS TR&T program.